# Helicity dependent photovoltaic effect in Bi$_2$Se$_3$ under normal incident light

*Jean Besbas, Karan Banerjee, Jaesung Son, Yi Wang, Yang Wu, Matthew Brahlek, Nikesh Koirala, Jisoo Moon, Seongshik Oh and Hyunsoo Yang*[*]

Dr. J. Besbas, K. Banerjee, Dr. J. Son, Dr. Y. Wang, Y. Wu, Prof. H. Yang
Department of Electrical and Computer Engineering, NUSNNI, National University of Singapore, 117576, Singapore
E-mail: eleyang@nus.edu.sg

M. Brahlek, N. Koirala, J. Moon, Prof. S. Oh
Department of Physics & Astronomy, Rutgers, The State University of New Jersey, Piscataway, New Jersey, 08854, USA

Keywords: topological insulators, Bi$_2$Se$_3$, photocurrent, circular photogalvanic effect, photogalvanic effect

Topological insulators (TIs) form a new class of materials with insulating bulk and surface conduction ensured by topologically protected surface states (TPSS). We investigate the impact of the helicity of a normally incident laser beam on the photovoltaic effect in the TI Bi$_2$Se$_3$. The observation of a helicity dependent photovoltaic effect for normally incident light indicates the presence of out-of-plane spin components for some TPSSs due to the hexagonal warping. In addition, fluctuations in the electrostatic potential at the surface locally break the rotational symmetry of the film allowing the helicity dependent photovoltaic effect. Our result suggests that engineering local electrostatic potentials in Bi$_2$Se$_3$ would allow the control of optically generated spin currents, which may be useful for applications in spin-optoelectronics.

## 1. Introduction

Recently discovered TIs have drawn significant attention for their intriguing properties.[1] A landscape of exotic optical phenomena are emerging in TI physics such as Majorana fermions for quantum optics,[2] realization of axion electrodynamics,[3, 4] universal Faraday effect[5] or Floquet states.[6] Among these, interesting optoelectronic effects are photogalvanic and photovoltaic effects, where light illumination results in the appearance of an electric current or voltage.[7] Before the demonstration of their topological properties, Bismuth based TIs, such as $Bi_2Se_3$ or $Bi_2Te_3$ were already known for their high thermoelectric power. Recent developments show that specific topological properties of TIs, relying on their topologically protected surface states (TPSSs), can be implemented in low cost photodetectors. Such properties include the optical control of surface electron scattering leading to an anomalous photoelectric effect in $Bi_2Te_3$,[8] the possibility of harvesting light from the ultraviolet to the terahertz range in $Bi_2Te_3$-Si composite photodetectors,[9] the photo thermoelectric effect on $Bi_2Te_3$ and $Sb_2Te_3$ surface steps,[10] or polarization dependent photovoltaic and photogalvanic effects in $Bi_2Se_3$, $Bi_2Te_3$ or $(Bi_{(1-x)}Sb_x)_2Te_3$ for polarization sensitive detection.[11-14]

Phenomenologically the total photovoltage ($V_{tot} = V_0 + V_L + V_C$) is the sum of a term $V_0$ independent of the applied light polarization, a term $V_L$, which depends on the direction of the linear polarization of the incident light, and a helicity dependent photovoltage (HDP) term $V_C$, which depends on the helicity of the incident light.[8, 13-18] In the prototype TI $Bi_2Se_3$, circular photogalvanic effect (CPGE) is one kind of helicity dependent effect, which contributes to $V_C$.

CPGE is the phenomenon wherein circularly polarized light incident onto a surface generates a spin polarized photocurrent.[8, 13-23] For obliquely incident light, the direction of the photocurrent ($j_x$) is along the axis perpendicular to the plane of the incidence, as shown in **Figure**

**1**a. Thus, the current and spin direction generated by CPGE can be manipulated by the angle and polarization of the incident light, leading to a full optical control of the spin photocurrent, which can be utilized for spintronics applications.[8, 13-18] Microscopically, the CPGE requires strong spin-orbit coupling and spin-momentum locking, i.e. the locking of the spin perpendicularly to the Bloch *k*-vector. Upon illumination with circularly polarized light, the spin of the charge carriers in the TIs is oriented through spin-orbit coupling, which transfers angular momentum from the light photons to the carrier spin momentum.[23, 24] Spin-momentum locking is a property of the TPSS[25-29] and the Rashba spin-split states[30-32] in $Bi_2Se_3$, which ensures that charge carriers with opposite spins propagate in opposite directions. It explains why the CPGE in $Bi_2Se_3$ has previously been attributed to the existence of the TPSS and cannot occur for normally incident light, because of the in-plane spin texture of TPSS near the Dirac point.[8, 13, 15-18] Indeed, with an in-plane spin texture, the angular momentum conservation forbids an optically induced transfer of angular momentum for normally incident light, as shown in **Figure 2**a. However, a HDP is theoretically possible even for normally incident light when there is an out-of-plane spin component for certain states in the spin texture. Therefore, the observation of HDP for normally incident light could indicate the existence of an out-of-plane spin texture in a TI.

In this work, we report the observation of HDP for normally incident light, indicating the existence of an out-of-plane spin component in the surface states of $Bi_2Se_3$, whose origin is the hexagonal warping of the Dirac cone away from the Dirac point.[33-39] An interpretation is proposed that could explain the observed helicity dependent photovoltaic effect due to a photocurrent caused by the interplay between hexagonal warping and local variations in the electrostatic potential in the $Bi_2Se_3$ film.

## 2. Sample Preparation and Photocurrent Measurements

Molecular beam epitaxy (MBE) implementing two-step scheme was used to obtain high quality 10 and 20 quintuple layer (QL, 1 QL ~ 1 nm) $Bi_2Se_3$ thin films grown on top of a c-axis $Al_2O_3$ substrate.[40, 41] For photovoltage measurements, the $Bi_2Se_3$ films were patterned using photo-lithography and Ar ion milling. The device pattern was designed to measure the Hall resistance and photocurrent responses. Ta/Cu/Ru (5/150/5 nm) or Ta/Cu (4/80 nm) electrode was deposited using sputtering on the 10 nm or 20 nm thick $Bi_2Se_3$ device, respectively. The 10 nm thick $Bi_2Se_3$ device was capped by $MgO/SiO_2$ (1/3 nm). A sample of different quality and origin based on a flake of $Bi_2Se_3$ was also investigated, which was exfoliated on an n-Si/$SiO_2$ substrate from polycrystalline $Bi_2Se_3$ from *Alfa Aesar*. The thickness of the flake was determined by atomic force microscopy (AFM) to be ~ 25 nm, which is comparable to the light penetration depth of 25 nm in such materials. Cr/Au (10/100 nm) contacts were subsequently patterned by lithography and deposited on the flake by thermal evaporation.[42] AFM images of 10 nm and 20 nm thick MBE $Bi_2Se_3$ films and the $Bi_2Se_3$ flake are presented in Supporting Figure S1.

As a control experiment, we demonstrate the HDP due to the CPGE in a 10 nm thick MBE $Bi_2Se_3$ device. In this experiment, a 650 nm semiconducting laser light with a power of ~ 25 mW is focused by a 20× microscope objective lens on the sample with a spot size of ~ 10 µm and at an incident angle of 45º. Electrodes were aligned perpendicularly with the plane of incidence of light as required to observe the CPGE as shown in Figure 1a. Figure 1b displays the effect of the modulation of light helicity on the photovoltage. We found a strong dependency of the HDP with light helicity leading to a change of sign in the photovoltage. In this case, the HDP was attributed to the CPGE and shows an important contribution of the TPSSs which was expected in this configuration.[13, 14, 17] An AFM profile of the 10 nm MBE film is presented in Figure 1c showing a low roughness of the film (see also Supporting Figure S1a).

In order to investigate the effect of the helicity of the normally incident light on the photovoltage, we developed a novel scanning microscope. For TIs, previous experimental approaches used scanning photocurrent microscopy to highlight local charge fluctuations on the surface, without resolving the influence of the polarization of light on the photocurrent generation.[43, 44] Moreover, most studies of the CPGE investigate the response to light polarization on a device area defined by the diameter of the laser spot.[13, 22] In $Bi_2Se_3$, this approach is useful to decrease a background that is attributed to the high thermoelectric power of Bi based TIs,[45-47] however it limits the choice of locations and detailed understanding of the relationship between effects depending on the light polarization. In order to circumvent this issue, we utilize a scanning photocurrent microscope to image the photovoltage and its dependency with light polarization in any device. Our method efficiently extracts the response of the photovoltage to the light helicity even in the presence of a helicity independent background. Our setup provides a spatial resolution to the HDP ($V_C$) and a direct comparison with the corresponding total photovoltage ($V_{tot}$) at the same location.

For the scanning experiment, a HeNe laser with a wavelength of 632.8 nm (1.96 eV) was used as a light source. The laser was focused on the sample using a 100× microscope objective lens as displayed in Figure 2b. The laser spot is seen in Figure 2c at the center of the $Bi_2Se_3$ film. The size of the spot is close to the limit of diffraction (~ 1 μm) and the power is ~ 0.5 mW. The sample was mounted on a piezo-stage that allows translational motion in all three directions. The generated photovoltage was mapped by moving the sample in the focal plane of the objective lens. Our configuration where the sample is moved and the beam is maintained at the same position ensures that the incident angle of the light remains constant during the scan. The constant incident angle is important since the amplitude of the CPGE, which is absent in normal incidence, depends on the incident angle of the light as discussed previously. To measure the total photovoltage, the intensity of the laser was modulated by a mechanical chopper at a frequency $f_{ch} < 800$ Hz.

When the chopper is used, the modulation of the photovoltage is given by $V_{tot}(I_0)\cos(\omega_{ch}t)$, where $I_0$ is the total light intensity (see Supporting Information S2). As displayed in Figure 2b, a lock-in amplifier is used to obtain $V_{tot}(I_0)$. In the first approximation, the measured voltage ($V$) is proportional to the photocurrent ($I$) through $V = RI$, where $R$ (~ 2.6 k$\Omega$ for MBE Bi$_2$Se$_3$ film) is the sum of the resistances of the film and contacts. In order to measure the HDP, the helicity of the laser light was modulated instead of its intensity. The modulation of helicity was carried out by replacing the mechanical chopper by a photoelastic modulator (PEM) as displayed in Figure 2b. The PEM is widely used to modulate the light polarization in various polarization related measurements.[48] To our knowledge, it is the first time that a PEM is used as a part of a scanning photovoltage microscope. The modulation of the polarization consists of a sinusoidal variation from left to right circular polarization at the constant frequency of the PEM. When applied to a device sensitive to light helicity, the polarization modulated light induces a periodic HDP in the device. The magnitude of the modulated HDP is proportional to the difference between the dc photovoltages as measured with a static illumination by a left and right circular polarization light. The PEM main axis is at 45º with respect to the linear polarization of the incoming light. In this configuration, the PEM acts on the polarization of the laser light as a rotating quarter waveplate. The intensity of the light is kept constant during the experiment whereas the polarization of the light is modulated with a fixed frequency $f_{PEM} \approx 50$ kHz. As shown in Figure 2b, the photovoltage is demodulated by using a lock-in amplifier, with the signal from the PEM controller acting as a reference for the lock-in amplifier. The modulated polarization dependent photovoltages are given by

$$\begin{cases} V_C(t) \propto \sqrt{2} I_0 i \left( \beta^l_{yx} - \beta^l_{xy} \right) J_1\left(\frac{\pi}{2}\right) \cos(\omega_{PEM} t) \\ V_L(t) \propto \frac{I_0}{\sqrt{2}} \left( \beta^l_{yx} + \beta^l_{xy} \right) \left( 1 + \left\{ J_0\left(\frac{\pi}{2}\right) - 2 J_2\left(\frac{\pi}{2}\right) \cos(2\omega_{PEM} t) \right\} \right) \end{cases} \quad (1)$$

where $J_n(x)$, $n = 0, 1, 2$ are Bessel functions of the first kind and $\beta^i_{jk}$ are components of the photogalvanic tensor (see Supporting Information S2). As seen from Equation (1), the first harmonic is attributed to the HDP and the second harmonic is attributed to a photovoltage depending on the light linear polarization. An optical image of the MBE $Bi_2Se_3$ film is shown in Figure 2c. Symbols (+ or −) indicate the connection toward the electrodes for the measurements. In Figure 2c, the blue box highlights the scanned area that can be directly observed on the reflectivity signal in Figure 2d. An AFM profile of the 20 nm MBE film is displayed in Figure 2e. The scanned area is limited by the range of the piezo stage.

## 3. Experimental Results

**Figure 3**a is a map of the total photovoltage ($V_{tot}$) when a normally incident laser shines at different locations of the device. The comparison between the reflectivity map in Figure 2d and the photovoltage map in Figure 3a allows an accurate determination of the device location where the photovoltaic effect is most efficient. Figure 3a shows that a photovoltage occurs mainly when light shines at the overlapping area between the electrode and the $Bi_2Se_3$ film. The predominant

mechanism for the emergence of the photovoltage at the metal/Bi$_2$Se$_3$ overlapping area is the Seebeck effect, wherein the temperature gradient $\Delta T$ across the electrode exposed to the laser and the un-exposed electrode gives rise to a potential difference $\Delta V = -S\Delta T$, where S is the Seebeck coefficient.[13] As discussed previously, a strong Seebeck effect is expected in Bi$_2$Se$_3$ due to a large thermoelectric effect with a Seebeck coefficient of S ~ $-100$ μV K$^{-1}$ (Seebeck coefficient of the metallic electrode can be neglected being ~1 μV K$^{-1}$).[45-47]

In Figure 3a, we observe that the photovoltage also appears when the laser shines on various locations on the bare Bi$_2$Se$_3$ film away from the electrode. The contribution from the bare film is about ten times weaker than the maximum photovoltage signal from the metal/Bi$_2$Se$_3$ overlapping area. The photovoltage in the bare film can be attributed to electron-hole photogeneration due to fluctuations in the local electrostatic potential, as recently reported.[43, 44] Furthermore, measurements using the scanning tunneling microscopy technique have demonstrated that the random distribution of defects in the Bi$_2$Se$_3$ bulk modifies the electric potential at the vicinity of the surface. As a consequence, the Dirac point of the TPSS may be slightly shifted at different positions in the film resulting in charge displacements.[49] Thus, the Bi$_2$Se$_3$ film is not perfectly homogenous, but consists of defect-induced charge puddles resulting in local n-n' junctions,[49] which can explain the observed photovoltage from the bare Bi$_2$Se$_3$ film (see Supporting Information S3).

Figure 3b shows a map of the HDP ($V_C$) obtained by modulating the helicity of the incident light with the PEM instead of the chopper. We first compare the total photovoltage and HDP signals at the metal/Bi$_2$Se$_3$ overlapping area, where the total photovoltage was found to be strongest in Figure 3a. At the same position, the HDP in Figure 3b is almost non-existent. Indeed, the HDP signal is strongest at the metal/Bi$_2$Se$_3$ interface and not in the overlapping region, as displayed in Figure 3b. This indicates the existence of polarization independent contributions ($V_0 \neq 0$) at the overlapping region and polarization dependent contributions ($V_C \neq 0$) at the interfacial region. The absence of HDP at the overlapping region confirms the purely thermal origin of the photovoltage at this position. Indeed, the Seebeck effect does not contribute to the HDP, since a thermal effect would contribute equally to the photocurrent on both left and right circularly polarized light.[50] A striking feature is that the HDP map in Figure 3b is quite similar to the total photovoltage map shown in Figure 3a in the area of the bare film. The similarity between the total photovoltage and HDP signals in the bare film indicates that they could share a common origin. As discussed above, local electrostatic fluctuations in the $Bi_2Se_3$ film are responsible for the photovoltage. Furthermore, the measurement of $V_L$ by demodulating the signal at the frequency of $2f_{PEM}$ did not produce any measurable signal, confirming $V_L \approx 0$ in agreement with the HDP presented in Figure 1b. Therefore, we conclude that $V_C$ contributes significantly to $V_{tot}$ in the location of the bare film. Consequently, both the total photovoltage and HDP could originate from local electrostatic potential fluctuations on the film surface.

**Figure 4** displays similar measurements on a $Bi_2Se_3$ flake exfoliated from $Bi_2Se_3$ from Alfa Aesar. Here, the distance between the electrodes is smaller (~ 5 µm) than that of the MBE film. Such a distance between electrodes in Figure 4a is comparable to the characteristic length scale of the temperature profile at the overlapping area between the electrode and $Bi_2Se_3$. Therefore, the total photovoltage between the electrodes in Figure 4b is attributed to the Seebeck effect, whose average profile between the electrodes is given in Figure 4c. A comparison of the profiles of the total photovoltage in Figure 4c and HDP in Figure 4e shows an absence of HDP at the location where the total photovoltage is the most important and confirms that the origin of the total photovoltage is mostly the Seebeck effect. Furthermore, the total photovoltage profile is in good agreement with a previous report and is caused by the temperature difference between the contacts as we observe in Figure 3a.[13] Instead, HDP measurements in Figure 4d,e display a weak but clearly identifiable HDP at the metal/$Bi_2Se_3$ interface, thus emphasizing the role of the interfacial electrostatic potential, similar to the data in Figure 3b. Therefore, we conclude that our observation of HDP for normally incident light is not restricted to a specific material preparation method, but can be universal. The surface roughness is similar to that of the MBE films as shown in Figure 4f.

We note that the microscopic origin of the photovoltaic effect, such as the dimensionality of the states that are involved, cannot be determined from scanning photovoltage microscopy alone. For instance, the generation of photocurrent by the photovoltaic effect in a p-n junction involves 3D bulk states in most semiconductors. However, graphene monolayers with 2D Dirac states are also known to cause a photocurrent generation in the vicinity of p-n junctions.[51-53] In order to further understand the nature of the states involved in the HDP, we performed time resolved magneto-optic Kerr effect (TR-MOKE) measurements on a MBE $Bi_2Se_3$ thin film (see Supporting Information S4). In this experiment, a circularly polarized femtosecond optical pulse with a wavelength of 800 nm illuminated the sample at normal incidence. In $Bi_2Se_3$, the circularly polarized pump excites out-of-plane spin polarized bulk electrons which have a characteristic spin decay time of ~ 0.2 ps.[54] Spin polarized surface electrons in TPSS with spins canted out-of-plane are excited by the pump as well. The spin polarization of the excited electrons is measured by a delayed normally incident probe pulse with a wavelength of 400 nm using the MOKE.

**Figure 5**a displays two measurements of the TR-MOKE signal, where the helicities of the pump beams were opposite. A common background signal between the two measurements results from a residual reflectivity dynamics. Both signals share a small oscillatory component whose frequency obtained by fast Fourier transform is 2.13 THz. These oscillations are known as coherent vibrations of the $A_{1g}$ longitudinal optical phonons of $Bi_2Se_3$ caused by the ultrafast cooling of bulk electrons after their excitation and thermalization.[55-58] Figure 5b shows the subtracted signal between the two measurements in Figure 5a. The subtracted signal is a direct measurement of the transient out-of-plane spin polarization induced in the film by the polarization of the pump beam. The pump induced spin polarization shows a decaying signal that can be fitted by two exponential functions. Keeping the smallest characteristic time at 0.2 ps, which is the reported spin lifetime for both bulk and surface states, the second spin relaxation time is determined at ~ 1.9 ps, which is between the bulk (~ 2.3 ps) and surface (~ 1.2 ps) intraband carriers cooling time through electron-phonon scattering.[56-58] This result is expected for a film of 20 nm where hot bulk electrons are partially trapped by the surface on a length defined by the sum of the Thomas-Fermi lengths of both surfaces (~ 10 nm).[57, 58] The strong CPGE displayed in Figure 1b further attests the important coupling of the TPSSs with light in the case of thin $Bi_2Se_3$ layer.

Though bulk properties cannot straightforwardly explain the dynamics, a natural explanation arises in term of surface electrons in the TPSS. Indeed, right after the initial thermalization, excited surface electrons with high temperature occupy states in the vicinity of the Fermi level which have in-plane spins in the low doping regime.[57, 58] Therefore, spins of hot surface electrons in the TPSS must be in-plane in the low doping case. On the contrary, significant hexagonal warping in the vicinity of the Fermi level in highly n-doped samples can play a role to prevent the relaxation of out-of-plane spin polarized surface electrons, and explain the correlation between cooling time and out-of-plane spin polarization lifetime. Our Hall effect measurements performed in $Bi_2Se_3$ MBE film (see Supporting Figure S2) yields a carrier density in agreement with the Fermi level lying in the conduction band (typically ~ 450 meV above the Dirac point), where moderate hexagonal warping has been reported for both TPSS and Rashba states.[33-39]

## 4. Discussion

The enhancement of out-of-plane spin lifetime of surface electrons suggests that HDP arises from the TPSS and/or the Rashba states in the $Bi_2Se_3$. The relative contribution of the TPSS and Rashba states is roughly estimated by comparing the amplitude of spin splitting $\hbar v_F/\alpha \approx 60$, where $v_F \approx 5\times10^5$ m/s is the Fermi velocity[59] and $\alpha \approx 0.36$ eV/s is the Rashba coupling.[30-32] The large magnitude of this ratio is in favor of a dominant contribution from the TPSS to the HDP.

The incident photon energies in our photovoltage (~ 1.96 eV) and TR-MOKE experiments (~ 1.55 eV) suggest several possible interband optical transitions involving TPSS. From band structure considerations, both optical transitions from bands below the $Bi_2Se_3$ valence band to the TPSS and from the TPSS to bands above the conduction band are possible. Moreover, recent reports discuss the possibility of optical transitions from TPSS to empty TPSS with higher energies.[60] However, optical selection rules in the electric dipole approximation forbid transitions between quantum states of identical symmetry. TPSS, bulk states and empty TPSS above the conduction band consist of orbitals with *p* symmetry, whereas bulk states below the valence band are of *s* symmetry. Therefore, it is more likely that visible light promotes electrons of *s* symmetry from bulk states below the valence band to the TPSS of *p* symmetry as shown in **Figure 6**a.[37]

Furthermore, the existence of a HDP and out-of-plane transient spin polarization under normally incident illumination suggest the presence of out-of-plane spin components in the spin texture, which can be accounted for by the hexagonal warping.[33-37] Figure 6a-b displays the mechanism of photogeneration in the band structure of $Bi_2Se_3$ with hexagonal warping. In Figure 6a-b, LCP (RCP) light generates spin down (up) electrons propagating in the direction of the wavevector at three distinct areas represented by the filled blue (textured red) triangles inside the hexagonal energy contour in Figure 6b. At these areas, the spin of the TPSS represented by arrows outside the hexagon, alternates from in-plane to partially out-of-plane in a three folded manner allowing optical excitation of specific TPSSs.

In general, the photogenerated currents from the above three areas cancel out each other, thus leading to an absence of total current at the $Bi_2Se_3$ surface under normally incident light excitation. However, our results suggest that the photogeneration on an electrostatic potential gradient leads to a spin dependent photocurrent. To discuss this point, one electrostatic step arbitrarily taken along the x axis is represented in reciprocal space in Figure 6b and in real space in Figure 6c. Let us consider two electrons with opposite spins excited on the electrostatic step at ~$(0, \pm k_F)$ occupying states in the areas depicted by black asterisks in Figure 6b and represented in real space in Figure 6c. In Figure 6c, the spin up electron created by RCP light will be reflected on the electrostatic step, whereas the spin down electron created by LCP light propagates away from the potential step. Moreover, the reflectivity coefficient of the spin up electrons on the electrostatic step is determined by the coupling between TPSSs at ~$(0, \pm k_F)$ represented by dashed arrows in Figure 6b. Similarly, some spin down electrons excited by LCP light at ~$k_F(-\cos(\pi/3), \pm \sin(\pi/3))$ will be also reflected by the step with a different reflectivity coefficient defined by the coupling between TPSSs at ~$k_F(-\cos(\pi/3), \pm\sin(\pi/3))$ and at ~$k_F(\cos(\pi/3), \pm\sin(\pi/3))$ represented by full arrows in Figure 6b. The asymmetry between reflectivities of spin up and spin down hot electrons on the electrostatic potential step results in different magnitudes of photocurrent for RCP and LCP light, thus explaining the HDP that we observed in $Bi_2Se_3$.

## 5. Conclusion

We have observed a HDP for normally incident light in TI $Bi_2Se_3$. HDP and spin lifetime measurements for normally incident light indicate the existence of out-of-plane spin components in the spin texture, that are attributed to the hexagonal warping of the TPSSs at the vicinity of the Fermi level. The helicity dependent photovoltaic effect is explained by a combination of hexagonal warping and local electrostatic potential gradients, resulting in spin dependent reflection of the spin polarized photogenerated electrons on the electrostatic fluctuation. Our result suggests the possibility of a spin dependent photogeneration at p-n junctions of TIs that could find applications in optoelectronics.

**Supporting Information**

Supporting Information is available from the Wiley Online Library or from the author.

**Acknowledgements**

This work was partially supported by the A*STAR's Pharos Programme on Topological Insulators, Ministry of Education-Singapore Academic Research Fund Tier 1 (R-263-000-B47-112) & Tier 2 (R-263-000-B10-112), Office of Naval Research (N000141210456), and by the Gordon and Betty Moore Foundation's EPiQS Initiative through Grant GBMF4418.

[1] M. Z. Hasan, C. L. Kane, *Rev. Mod. Phys.* **2010**, 82, 3045.
[2] L. Fu, C. L. Kane, *Phys. Rev. Lett.* **2008**, 100, 096407.
[3] X.-L. Qi, T. L. Hughes, S.-C. Zhang, *Phys. Rev. B* **2008**, 78, 195424.
[4] J. Maciejko, X.-L. Qi, H. D. Drew, S.-C. Zhang, *Phys. Rev. Lett.* **2010**, 105, 166803.
[5] R. Valdés Aguilar, A. V. Stier, W. Liu, L. S. Bilbro, D. K. George, N. Bansal, L. Wu, J. Cerne, A. G. Markelz, S. Oh, N. P. Armitage, *Phys. Rev. Lett.* **2012**, 108, 087403.
[6] Y. H. Wang, H. Steinberg, P. Jarillo-Herrero, N. Gedik, *Science* **2013**, 342, 453.
[7] V. I. Belinicher, B. I. Sturman, *Phys. -Usp.* **1980**, 23, 199.
[8] H. Zhang, J. Yao, J. Shao, H. Li, S. Li, D. Bao, C. Wang, G. Yang, *Sci. Rep.* **2014**, 4, 5876.
[9] J. Yao, J. Shao, Y. Wang, Z. Zhao, G. Yang, *Nanoscale* **2015**, 7, 12535.
[10] J. H. Sung, H. Heo, I. Hwang, M. Lim, D. Lee, K. Kang, H. C. Choi, J.-H. Park, S.-H. Jhi, M.-H. Jo, *Nano Lett.* **2014**, 14, 4030.
[11] K. N. Okada, N. Ogawa, R. Yoshimi, A. Tsukazaki, K. S. Takahashi, M. Kawasaki, Y. Tokura, *Phys. Rev. B* **2016**, 93, 081403(R).
[12] J. D. Yao, J. M. Shao, S. W. Li, D. H. Bao, G. W. Yang, *Sci. Rep.* **2015**, 5, 14184.
[13] J. W. McIver, D. Hsieh, H. Steinberg, P. Jarillo-Herrero, N. Gedik, *Nat. Nanotechnol.* **2012**, 7, 96.
[14] Y. Yan, Z.-M. Liao, X. Ke, G. Van Tendeloo, Q. Wang, D. Sun, W. Yao, S. Zhou, L. Zhang, H.-C. Wu, D.-P. Yu, *Nano Lett.* **2014**, 14, 4389.
[15] Q. S. Wu, S. N. Zhang, Z. Fang, X. Dai, *Phys. E* **2012**, 44, 895.
[16] A. Junck, G. Refael, F. von Oppen, *Phys. Rev. B* **2013**, 88, 075144.
[17] J. Duan, N. Tang, X. He, Y. Yan, S. Zhang, X. Qin, X. Wang, X. Yang, F. Xu, Y. Chen, W. Ge, B. Shen, *Sci. Rep.* **2014**, 4, 4889.
[18] C. Kastl, C. Karnetzky, H. Karl, A. W. Holleitner, *Nat. Commun.* **2015**, 6, 6617.
[19] B. Wittmann, L. E. Golub, S. N. Danilov, J. Karch, C. Reitmaier, Z. D. Kvon, N. Q. Vinh, A. F. G. van der Meer, B. Murdin, S. D. Ganichev, *Phys. Rev. B* **2008**, 78, 205435.
[20] B. Wittmann, S. N. Danilov, V. V. Bel'kov, S. A. Tarasenko, E. G. Novik, H. Buhmann, C. Brüne, L. W. Molenkamp, Z. D. Kvon, N. N. Mikhailov, S. A. Dvoretsky, N. Q. Vinh, A. F. G. van der Meer, B. Murdin, S. D. Ganichev, *Semicond. Sci. Technol.* **2010**, 25, 095005.
[21] J. Karch, P. Olbrich, M. Schmalzbauer, C. Zoth, C. Brinsteiner, M. Fehrenbacher, U. Wurstbauer, M. M. Glazov, S. A. Tarasenko, E. L. Ivchenko, D. Weiss, J. Eroms, R. Yakimova, S. Lara-Avila, S. Kubatkin, S. D. Ganichev, *Phys. Rev. Lett.* **2010**, 105, 227402.
[22] H. Yuan, X. Wang, B. Lian, H. Zhang, X. Fang, B. Shen, G. Xu, Y. Xu, S.-C. Zhang, H. Y. Hwang, Y. Cui, *Nat. Nanotechnol.* **2014**, 9, 851.
[23] S. D. Ganichev, W. Prettl, *J. Phys.: Condens. Matter* **2003**, 15, R935.
[24] *Optical Orientation* (Eds: F. Meier, B. P. Zakharchenya), Elsevier, Amsterdam, The Netherlands **1984**.
[25] H. Zhang, C.-X. Liu, X.-L. Qi, X. Dai, Z. Fang, S.-C. Zhang, *Nat. Phys.* **2009**, 5, 438.
[26] D. Hsieh, Y. Xia, D. Qian, L. Wray, F. Meier, J. H. Dil, J. Osterwalder, L. Patthey, A. V. Fedorov, H. Lin, A. Bansil, D. Grauer, Y. S. Hor, R. J. Cava, M. Z. Hasan, *Phys. Rev. Lett.* **2009**, 103, 146401.
[27] O. V. Yazyev, J. E. Moore, S. G. Louie, *Phys. Rev. Lett.* **2010**, 105, 266806.
[28] Z.-H. Pan, E. Vescovo, A. V. Fedorov, D. Gardner, Y. S. Lee, S. Chu, G. D. Gu, T. Valla, *Phys. Rev. Lett.* **2011**, 106, 257004.

[29]    C. Jozwiak, C.-H. Park, K. Gotlieb, C. Hwang, D.-H. Lee, S. G. Louie, J. D. Denlinger, C. R. Rotundu, R. J. Birgeneau, Z. Hussain, A. Lanzara, *Nat. Phys.* **2013**, 9, 293.
[30]    P. D. C. King, R. C. Hatch, M. Bianchi, R. Ovsyannikov, C. Lupulescu, G. Landolt, B. Slomski, J. H. Dil, D. Guan, J. L. Mi, E. D. L. Rienks, J. Fink, A. Lindblad, S. Svensson, S. Bao, G. Balakrishnan, B. B. Iversen, J. Osterwalder, W. Eberhardt, F. Baumberger, P. Hofmann, *Phys. Rev. Lett.* **2011**, 107, 096802.
[31]    Z.-H. Zhu, G. Levy, B. Ludbrook, C. N. Veenstra, J. A. Rosen, R. Comin, D. Wong, P. Dosanjh, A. Ubaldini, P. Syers, N. P. Butch, J. Paglione, I. S. Elfimov, A. Damascelli, *Phys. Rev. Lett.* **2011**, 107, 186405.
[32]    H. M. Benia, A. Yaresko, A. P. Schnyder, J. Henk, C. T. Lin, K. Kern, C. R. Ast, *Phys. Rev. B* **2013**, 88, 081103(R).
[33]    K. Kuroda, M. Arita, K. Miyamoto, M. Ye, J. Jiang, A. Kimura, E. E. Krasovskii, E. V. Chulkov, H. Iwasawa, T. Okuda, K. Shimada, Y. Ueda, H. Namatame, M. Taniguchi, *Phys. Rev. Lett.* **2010**, 105, 076802.
[34]    Y. H. Wang, D. Hsieh, D. Pilon, L. Fu, D. R. Gardner, Y. S. Lee, N. Gedik, *Phys. Rev. Lett.* **2011**, 107, 207602.
[35]    S. Souma, K. Kosaka, T. Sato, M. Komatsu, A. Takayama, T. Takahashi, M. Kriener, K. Segawa, A. Yoichi, *Phys. Rev. Lett.* **2011**, 106, 216803.
[36]    S. Basak, H. Lin, L. A. Wray, S.-Y. Xu, L. Fu, M. Z. Hasan, A. Bansil, *Phys. Rev. B* **2011**, 84, 121401(R).
[37]    S. R. Park, J. Han, C. Kim, Y. Y. Koh, C. Kim, H. Lee, H. J. Choi, J. H. Han, K. D. Lee, N. J. Hur, M. Arita, K. Shimada, H. Namatame, M. Taniguchi, *Phys. Rev. Lett.* **2012**, 108, 046805.
[38]    Y. Wang, N. Gedik, *Phys. Status Solidi RRL* **2013**, 7, 64.
[39]    Y. Wang, P. Deorani, K. Banerjee, N. Koirala, M. Brahlek, S. Oh, H. Yang, *Phys. Rev. Lett.* **2015**, 114, 257202.
[40]    N. Bansal, Y. S. Kim, E. Edrey, M. Brahlek, Y. Horibe, K. Iida, M. Tanimura, G.-H. Li, T. Feng, H.-D. Lee, T. Gustafsson, E. Andrei, S. Oh, *Thin Solid Films* **2011**, 520, 224.
[41]    N. Bansal, Y. S. Kim, M. Brahlek, E. Edrey, S. Oh, *Phys. Rev. Lett.* **2012**, 109, 116804.
[42]    J. Son, K. Banerjee, M. Brahlek, N. Koirala, S.-K. Lee, J.-H. Ahn, S. Oh, H. Yang, *Appl. Phys. Lett.* **2013**, 103, 213114.
[43]    C. Kastl, T. Guan, X. Y. He, K. H. Wu, Y. Q. Li, A. W. Holleitner, *Appl. Phys. Lett.* **2012**, 101, 251110.
[44]    C. Kastl, P. Seifert, X. He, K. Wu, Y. Li, A. Holleitner, *2D Mater.* **2015**, 2, 024012.
[45]    S. K. Mishra, S. Satpathy, O. Jepsen, *J. Phys.: Condens. Matter* **1997**, 9, 461.
[46]    A. Al Bayaz, A. Giani, A. Foucaran, F. Pascal-Delannoy, A. Boyer, *Thin Solid Films* **2003**, 441, 1.
[47]    G. Li, D. Liang, R. L. J. Qiu, X. P. A. Gao, *Appl. Phys. Lett.* **2013**, 102, 043104.
[48]    S. Polisetty, J. Scheffler, S. Sahoo, Y. Wang, T. Mukherjee, X. He, C. Binek, *Rev. Sci. Instrum.* **2008**, 79, 055107.
[49]    H. Beidenkopf, P. Roushan, J. Seo, L. Gorman, I. Drozdov, Y. S. Hor, R. J. Cava, A. Yazdani, *Nat. Phys.* **2011**, 7, 939.
[50]    S. D. Ganichev, H. Ketterl, W. Prettl, E. L. Ivchenko, L. E. Vorobjev, *Appl. Phys. Lett.* **2000**, 77, 3146.

[51]	F. Xia, T. Mueller, R. Golizadeh-Mojarad, M. Freitag, Y.-m. Lin, J. Tsang, V. Perebeinos, P. Avouris, *Nano Lett.* **2008**, 9, 1039.
[52]	E. C. Peters, E. J. H. Lee, M. Burghard, K. Kern, *Appl. Phys. Lett.* **2010**, 97, 193102.
[53]	G. Rao, M. Freitag, H.-Y. Chiu, R. S. Sundaram, P. Avouris, *ACS Nano* **2011**, 5, 5848.
[54]	D. Hsieh, F. Mahmood, J. W. McIver, D. R. Gardner, Y. S. Lee, N. Gedik, *Phys. Rev. Lett.* **2011**, 107, 077401.
[55]	Y. D. Glinka, S. Babakiray, T. A. Johnson, M. B. Holcomb, D. Lederman, *Appl. Phys. Lett.* **2015**, 117, 165703.
[56]	J. Qi, X. Chen, W. Yu, P. Cadden-Zimansky, D. Smirnov, N. H. Tolk, I. Miotkowski, H. Cao, Y. P. Chen, Y. Wu, S. Qiao, Z. Jiang, *Appl. Phys. Lett.* **2010**, 97, 182102.
[57]	Y. D. Glinka, S. Babakiray, J. T. A., M. B. Holcomb, D. Lederman, *Appl. Phys. Lett.* **2014**, 105, 171905.
[58]	Y. D. Glinka, S. Babakiray, T. A. Johnson, A. D. Bristow, M. B. Holcomb, D. Lederman, *Appl. Phys. Lett.* **2013**, 103, 151903.
[59]	Y. Xia, D. Qian, D. Hsieh, L. Wray, A. Pal, H. Lin, A. Bansil, D. Grauer, Y. S. Hor, R. J. Cava, M. Z. Hasan, *Nat. Phys.* **2009**, 5, 398.
[60]	J. A. Sobota, S.-L. Yang, A. F. Kemper, J. J. Lee, F. T. Schmitt, W. Li, R. G. Moore, J. G. Analytis, I. R. Fisher, P. S. Kirchmann, T. P. Devereaux, Z.-X. Shen, *Phys. Rev. Lett.* **2013**, 111, 136802.

**Figure 1.** CPGE in MBE grown $Bi_2Se_3$ with 10 nm thickness. a) Principle of photocurrent generation by circularly polarized light with oblique incidence using the CPGE. The laser is a semiconducting laser with a wavelength of 650 nm and power of 25 mW. The laser beam was modulated by a chopper with a linear polarizer and then passed through a quarter waveplate to control the circular polarization. The laser beam was focused on the sample using a 20× microscope objective lens and the voltage between the device electrodes was measured by a lock-in amplifier. b) Experimental total photovoltage from a $Bi_2Se_3$ sample illuminated with light at an incidence angle of $\theta \approx 45°$. The data were measured by rotating the quarter waveplate, which changed the polarization. Black circles are the experimental data and red line is a sine fit. c) Thickness profile of a 10 nm MBE film using AFM.

**Figure 2**. Principle of detection of helicity dependent photovoltage. a) With normally incident light, CPGE is not expected to occur. b) Sketch of the experimental setup. The laser beam is focused onto the sample after intensity or helicity modulation by the chopper or the PEM, respectively. Reflected light and photovoltage are detected simultaneously by the photodetector and the lock-in amplifier, respectively. c) Optical microscopy image of the sample. The orange dot in the middle is the attenuated laser spot with a spot size of ~ 1 μm and wavelength of 632.8 nm. The laser power is ~ 0.5 mW in the focal plane. Red symbols (+ or −) indicate the connection toward the electrodes. The blue square surrounds the scanned area. d) Scanning reflectivity map obtained with a photodetector in the scanned area. e) Thickness profile of a 20 nm MBE film using AFM.

**Figure 3.** Scanning photovoltage microscopy data from a MBE grown $Bi_2Se_3$ device with 20 nm thickness. Laser beam at a wavelength of 632.8 nm is focused at normal incidence by a 100×

microscope objective lens on a spot of diameter of ~ 1 µm. Photovoltage and HDP maps are obtained my moving a piezo stage in the focal plane of the objective lens. a) Total photovoltage map with intensity modulation by the chopper and linearly polarized light. b) HDP maps of the same area. HDP response is measured with light circular polarization modulation by the PEM.

**Figure 4.** Photovoltage maps from an exfoliated flake of $Bi_2Se_3$ from *Alfa Aesar*. Measurements are performed in the same conditions as for the MBE $Bi_2Se_3$ 20 nm thick film. a) Reflectivity map. b) Total photovoltage map between the Au electrodes with intensity modulation by a chopper and linearly polarized light. c) Averaged total photovoltage between electrodes from the data in b). Corresponding x-axis is shown at the left side of a). d) Map of the HDP after removing a constant background. The inset in d) shows a picture of the device. e) Averaged helicity dependent photovoltage between two electrodes calculated from d) Corresponding x-axis is displayed in d). f) AFM thickness profile of the flake surface.

**Figure 5.** Spin lifetime of carriers of a 20 nm thick MBE $Bi_2Se_3$ film obtained by TR-MOKE in polar configuration and at room temperature. Laser pulses at a repetition rate of 1 kHz are used. Pump excitation beam at wavelength ~ 800 nm and power ~ 1.2 mW is focused on a 300 µm spot. Probe beam at wavelength ~ 400 nm and power ~ 20 µW is focused on a 100 µm spot in the center of the pump spot. Polarization analysis of the Kerr rotation of the probe is performed using a balanced photodetector bridge. a) TR-MOKE dynamics triggered by optical femtosecond pulse excitation and left circularly polarized (LCP) or right circularly polarized (RCP) light. b) Decay of the optically generated spin polarization calculated from a). The black line shows the experimental data and the red line is a biexponential fit.

**Figure 6.** Mechanism of helicity dependent photovoltaic effect with normal incident light. a) Helicity dependent optical transitions involving TPSS for RCP and LCP light. b) Sketch of the energy contour and spin directions of the TPSS in the vicinity of the Fermi level in the presence of hexagonal warping. Triangles represent the states where spin-up (red textured) and spin-down (blue filled) electrons are photogenerated. Outer small arrows display the direction of the in-plane spin components. The vertical green line is one possible electrostatic potential step. Relevant scattering events of photogenerated electrons induced by the electrostatic potential step during the electron relaxation are represented by horizontal arrows in the hexagon. c) Spin dependent scattering of spin-polarized photogenerated electrons on the electrostatic potential step in real space for electrons with $k_y$ close to 0.

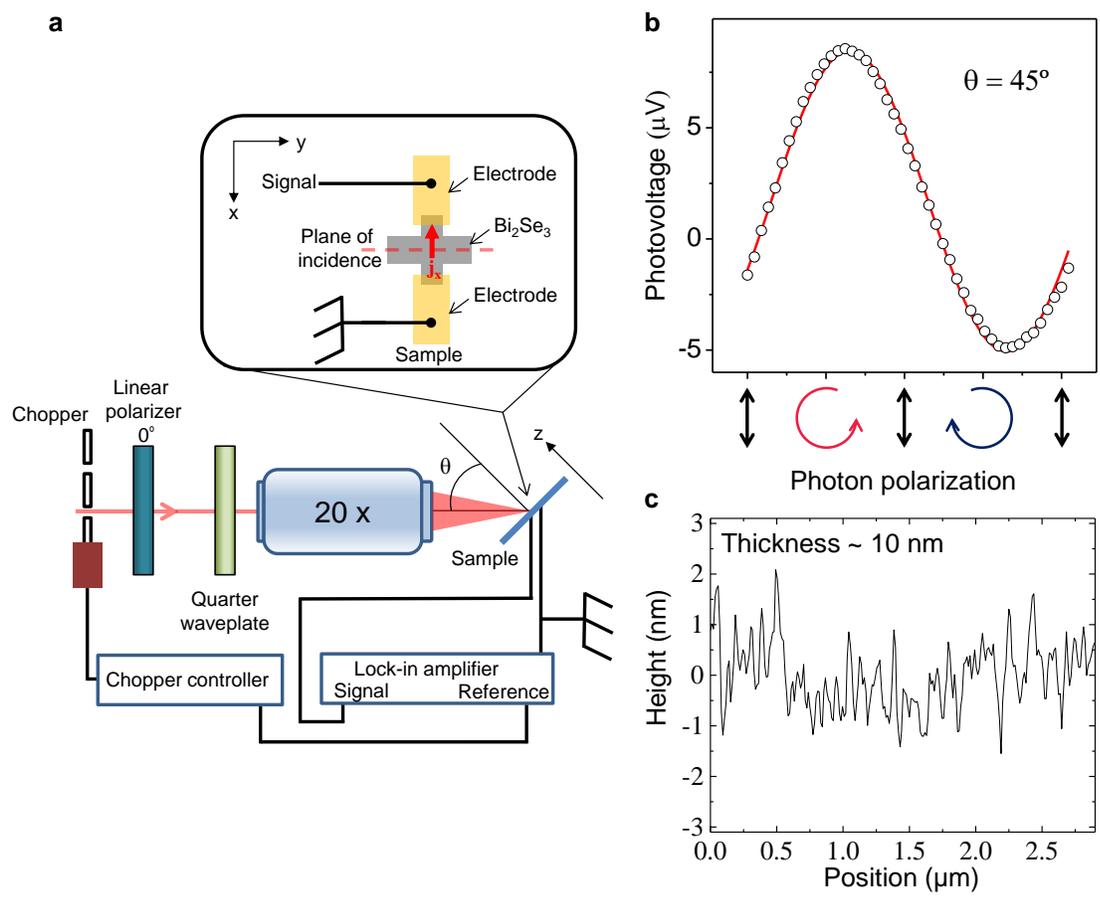

**Figure 1.**

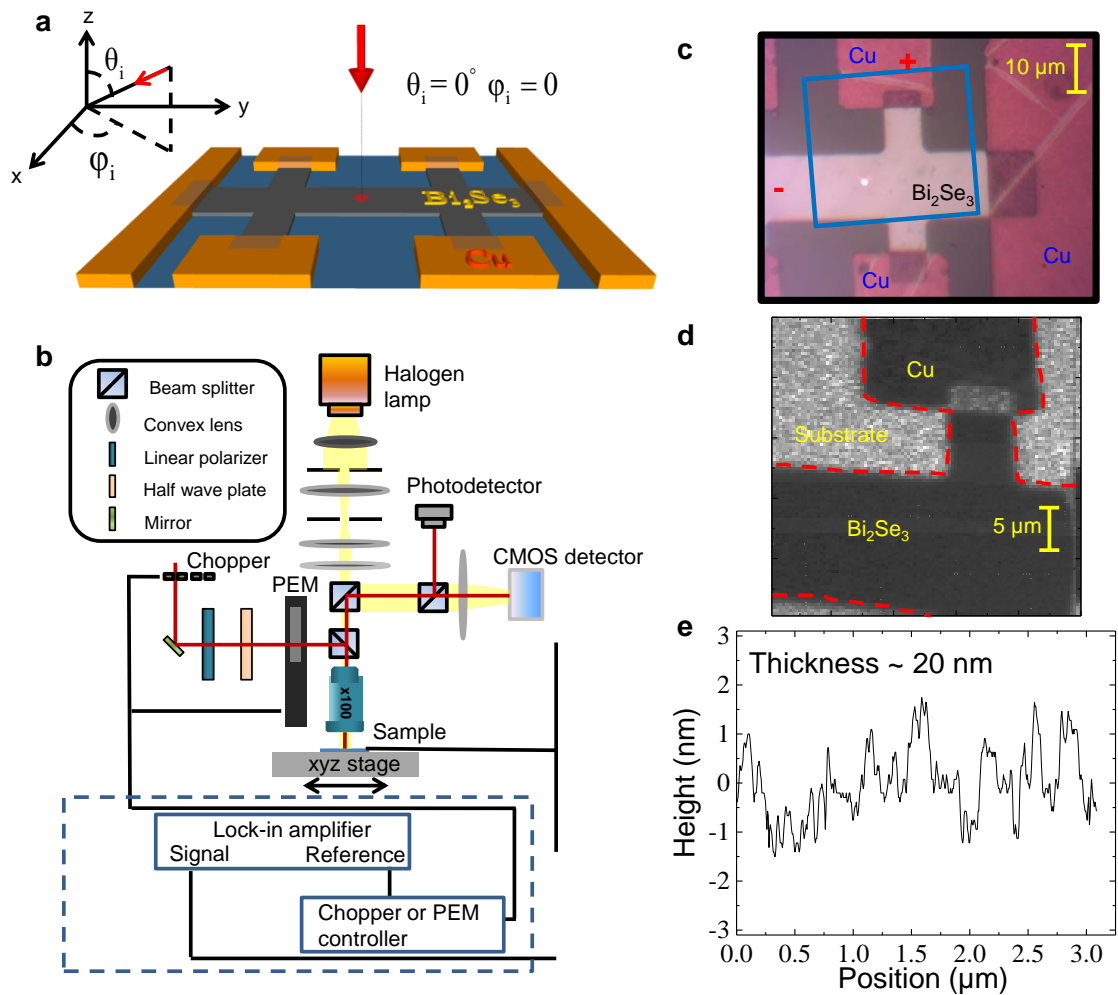

**Figure. 2**

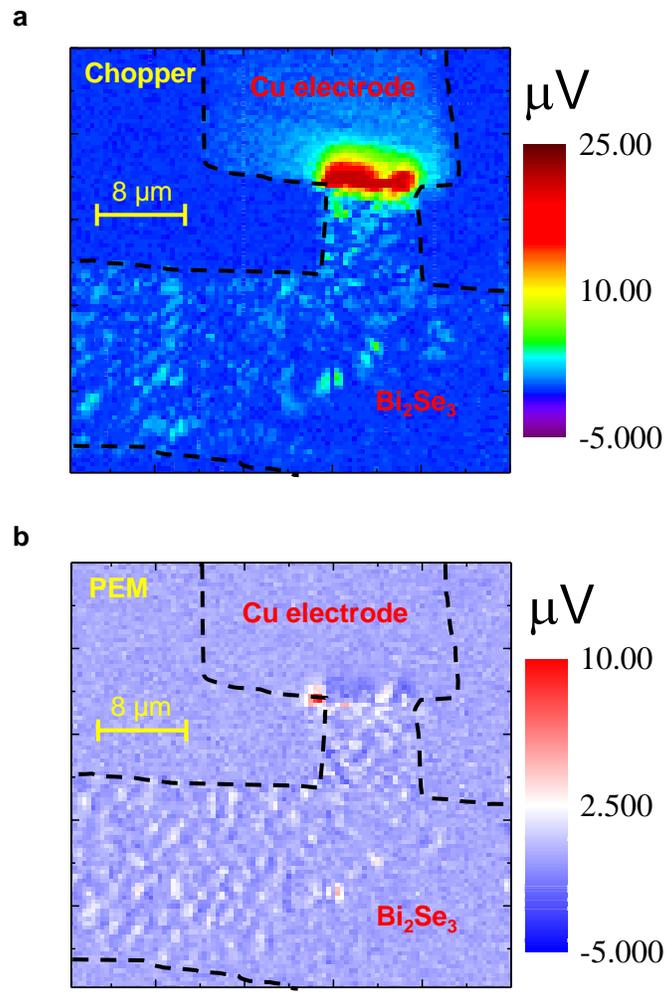

**Figure 3.**

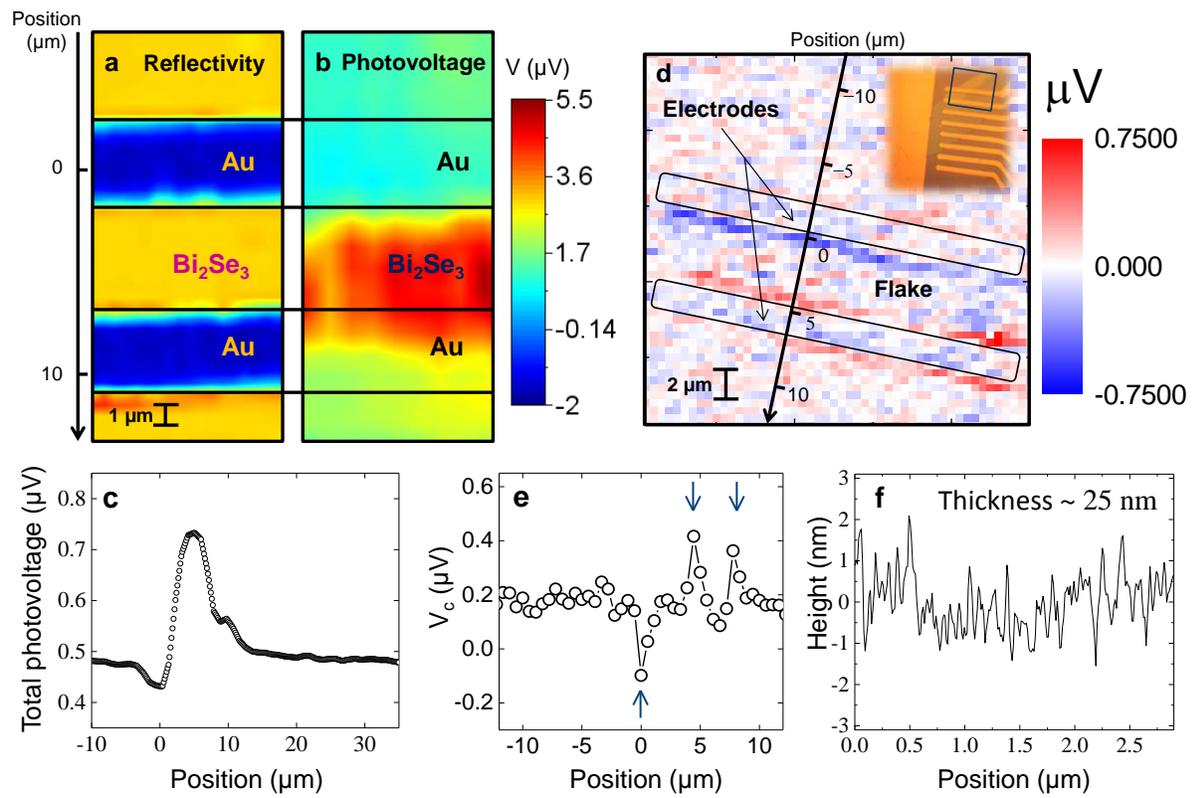

**Figure 4.**

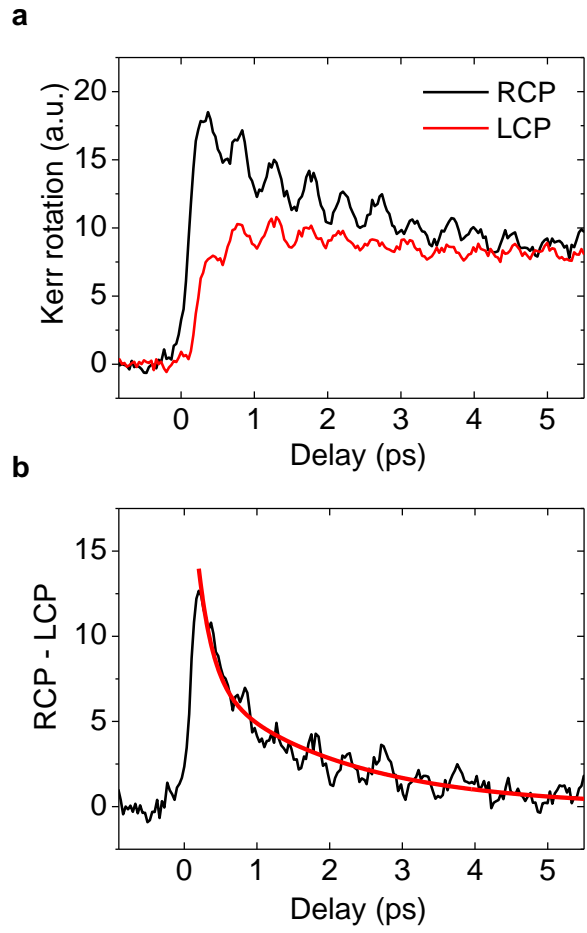

**Figure 5.**

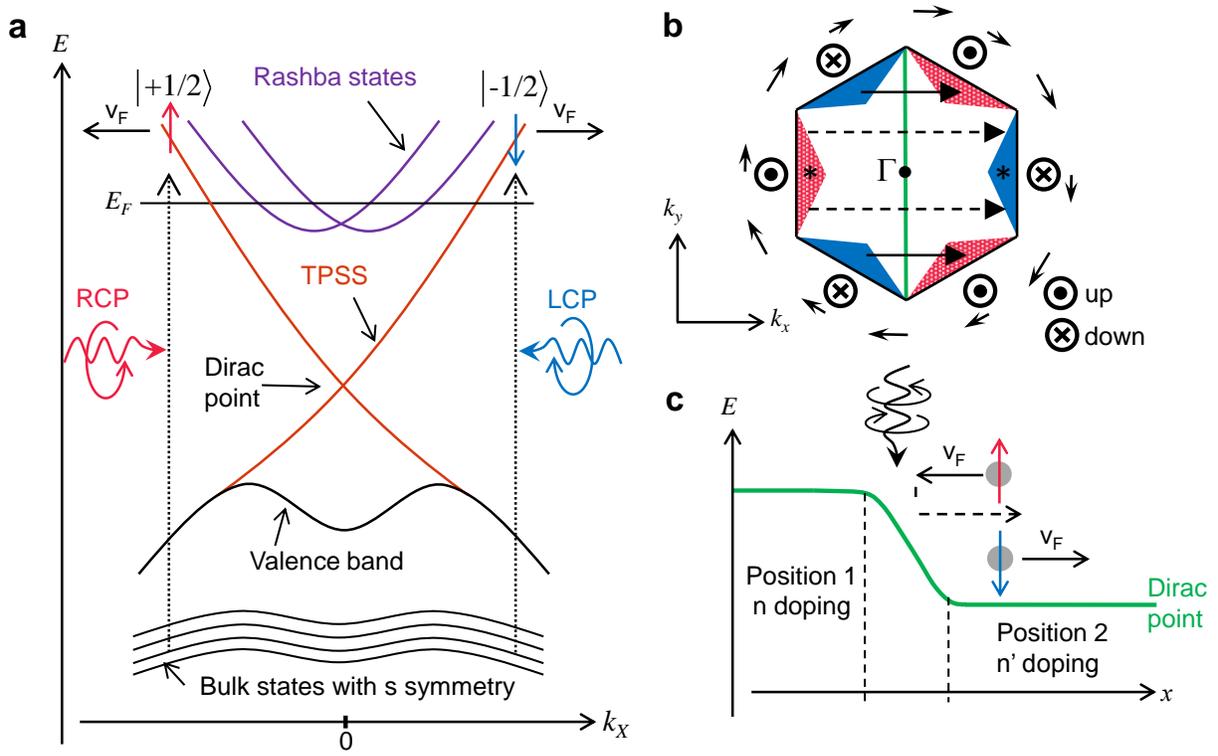

**Figure 6**.